\documentclass[preprint,aps,floats,nofootinbib,preprintnumbers]{revtex4}
\usepackage{graphicx}
\usepackage{amsmath}
\usepackage{amssymb}
\usepackage{bm}

\makeatletter
\@addtoreset{equation}{section}
\makeatother
\begin{document}
\preprint{KIAS-P15064}
\title{\Large Oblique corrections from less-Higgsless models\\ in warped space}
\author{Hisaki Hatanaka}
\affiliation{Quantum Universe Center, 
Korea Institute for Advanced Study,\\
Seoul 130-722, Republic of Korea}
\begin{abstract}\noindent
The Higgsless model in warped extra dimension is reexamined. Dirichlet boundary conditions on the TeV brane are replaced with Robin boundary conditions which are parameterized by a mass parameter $M$. We calculate the Peskin-Takeuchi precision parameters $S$, $T$ and $U$ at tree level. 
We find that to satisfy the constraints on the precision parameters at $99 \%$ [$95 \%$] confidence level (CL) the first Kaluza-Klein excited $Z$ boson, $Z'$, should be heavier than 5 TeV [8 TeV]. The Magnitude of $M$, which is infinitely large in the original model, should be smaller than 200 GeV (70 GeV) for the curvature of the warped space $R^{-1}=10^{16}$ GeV ($10^{8}$ GeV) at $95\%$ CL. If the Robin boundary conditions are induced by the mass terms localized on the TeV brane, from the $99\%$ [$95\%$] bound we find that the brane mass interactions account for more than $97\%$ [$99\%$] of the masses of $Z$ and $W$ bosons. Such a brane mass term is naturally interpreted as a vacuum expectation value of the Higgs scalar field in the standard model localized on the TeV brane. If so, the model can be tested by precise measurements of $HWW$, $HZZ$ couplings and search for 1st Kaluza-Klein excited states. 
\end{abstract}
\maketitle
\clearpage
\allowdisplaybreaks
\newcommand{\Tr}{{\mathop{\mbox{Tr}}\nolimits}} 
\newcommand{\tr}{{\mathop{\mbox{tr}}\nolimits}} 
\newcommand{\diag}{{\mathop{\mbox{diag}}\nolimits}} 
\newcommand{\calA}{{\cal A}}
\newcommand{\calL}{{\cal L}}\renewcommand{\L}{{\cal L}}
\newcommand{\calO}{{\cal O}}
\newcommand{\calX}{{\cal X}}
\newcommand{\rmA}{\text{A}}
\newcommand{\rmB}{\text{B}}
\newcommand{\tildeg}{\tilde{g}}
\newcommand{\GeV}{{\rm GeV}}
\newcommand{\TeV}{{\rm TeV}}
\section{Introduction}
Even after the discovery of the Higgs scalar with $125\GeV$ mass \cite{Aad:2012tfa,Chatrchyan:2012xdj},
mechanisms to maintain the hierarchy between the electroweak scale and Planck scale is still unknown.
Warped extra dimension is one of the way to explain such a large hierarchy between the electroweak scale and Planck scale \cite{Randall:1999ee}. In this scenario, such hierarchy is obtained from the exponentially large warp factor of the metric of the space.
In this direction, the standard model in the warped space is considered in \cite{Huber:2000fh}.
Such models, however, suffer from large deviation of oblique $S$ and $T$ parameters \cite{Peskin:1990zt,Peskin:1991sw}.
To suppress the $T$ parameter models are extended so as to possess the custodial
symmetry\cite{Agashe:2003zs}.
To suppress the $S$ parameter, brane-localized kinetic terms \cite{Davoudiasl:2002ua} and the soft-wall warped extra dimension are also considered in \cite{Cabrer:2011fb}.
Most cases, nevertheless, Kaluza-Klein (KK) scale is needed to be higher than $3\,\TeV$ to suppress the $S$ parameter. 

Although some excesses with invariant masses around or below $2\,\TeV$ in di-boson channels
have been reported \cite{Aad:2015owa,ATLAS15DB,Khachatryan:2014hpa,CMS:2015gla,ATLAS15DP,CMS:2015dxe}, 
experimental results in the LHC Run-1 \cite{ATLAS:2014wra,Khachatryan:2015pua,Aad:2014cka,Khachatryan:2014fba} and first-year results of LHC Run-2
\cite{1512.01224,CMS15W,CMS15Wtb,CMS15Z,ATLAS15W,ATLAS15Z}
seem to exclude the $Z'$ and $W'$ bosons which are lighter than $3\,\TeV$ in fermionic decay channels.
Therefore in this paper we focus on warped extra dimensional scenarios in which KK particles are heavier than $3\,\TeV$.

In this paper we reconsider the Higgsless model in warped extra dimension \cite{Csaki:2003zu,Cacciapaglia:2004jz,Cacciapaglia:2004rb}.
In the Higgsless model the electroweak symmetry breaking is caused by the boundary conditions on the TeV brane, and this model also yields large value of $S$ parameter\cite{Cacciapaglia:2004rb,Burdman:2003ya,Davoudiasl:2004pw}, and experimentally excluded by the discovery of the Higgs boson. In order to suppress the $S$ parameter, some Dirichlet boundary conditions on the brane are replaced with generalized Robin boundary conditions. A mass parameter $M$ is introduced to parameterize the Robin boundary conditions. In the $M\to\infty$ limit the model reduces to the original model, whereas $M=0$ reproduce the unbroken electroweak symmetry.
As $M$ decreases from $+\infty$ to zero, we obtain smaller magnitudes of $S$, $T$ and $U$ parameters while the Kaluza-Klein scale becomes larger. 

In this paper we also study the mass structure of weak bosons in detail.
The Robin boundary conditions can be induced by the mass terms localized on branes \cite{Gherghetta:2000qt,Barbieri:2003pr}.
In the original model where $M\to\infty$, the mass of weak bosons are coming from their momenta along the extra spacial dimension. As $M$ decreases, contributions from the brane mass terms dominates in the weak boson masses. Such a brane mass can also be identified with the vacuum expectation value (VEV) of a scalar field, namely the Higgs boson observed in the LHC. 
Based on such identification we also estimate the Higgs couplings to the weak bosons in this model.

This paper is organized as follows. In Section~\ref{sec:model}, an extension of the Higgsless model in warped space is introduced. In Section~\ref{sec:numeric}, the model is numerically studied. Section~\ref{sec:summary} is devoted to a summary and discussion. In Appendix~\ref{sec:wavefunc}, formulas for the wave function of the gauge field are collected.

\section{Model}\label{sec:model}
The model \cite{Csaki:2003zu,Cacciapaglia:2004jz} is a $SU(2)_L \otimes SU(2)_R \otimes U(1)_{B-L}$ gauge theory in a slice of five dimensional (5D) anti-de Sitter space $AdS_5$.
The metric of $AdS_5$ bulk is given by
\begin{eqnarray}
ds^2 &=& \frac{R^2}{z^2} [\eta_{MN} dx^M dx^N],
\quad
R \le z \le R',
\end{eqnarray}
where $M,N=0,1,2,3,5$, $\eta_{MN} = \diag(1,-1,-1,-1,-1)$ and $x^5 \equiv z$.
$R$ is the $AdS_5$ curvature radius.
A large hierarchy between $R$ and $R'$ appears as $\ln(R'/R) = \calO(10)$.
Boundaries at $z=R$ and $z=R'$ are referred as the Planck (UV) brane and the TeV (IR) brane, respectively.
Gauge fields propagate in $AdS_5$ bulk.
Let $A_M^{La}, A_M^{Ra}, B_M$  ($a=1,2,3$) be 5D gauge fields of $SU(2)_L$, $SU(2)_R$ and $U(1)_{B-L}$, respectively. 
The action of the gauge fidlds in the bulk is given by
\begin{eqnarray}
S_{\text{bulk}} &=& S_5[A^L] + S_5[A^R] + S_5[B],
\nonumber
\\
S_5[A] &\equiv& \int d^4 x \int_{R}^{R'} dz \frac{R}{z} \biggl\{
-\frac{1}{4} F_{\mu\nu}^a F^{a\mu\nu} 
+\frac{1}{2} (D_\mu A_5^a)^\dag(D^\mu A_5^a)
\nonumber\\&&
- \frac{1}{2\xi_A} \left[\partial_\mu A^{a\mu}
 - \xi_A z \partial_5 \left(\frac{1}{z}A_5^a \right) \right]^2
 \biggr\},
\end{eqnarray}
where $\mu,\nu=0,1,2,3$ and contractions of indices $\mu,\,\nu$ are done with $\eta_{\mu\nu}$.
$F_{\mu\nu}^a \equiv \partial_\mu A_\nu^a - \partial_\nu A_\mu^a + g_A f_{abc} A_\mu^b A_\nu^c$, and $D_\mu A_5^a = \partial_\mu A_5^a + g_A f_{abc} A_\mu^b A_5^c$. 
$f_{abc}$ is the structure constant of the gauge group, and vanishes for $U(1)_{B-L}$. 
$g_A = g_{5L},\,g_{5R},\,\tildeg_5$ denote the 5D gauge couplings of $SU(2)_L$, $SU(2)_R$ and $U(1)_{B-L}$. Hereafter we impose $SO(4)\simeq SU(2)_R \otimes SU(2)_L$ symmetry and 
set $g_{5L} = g_{5R} \equiv g_5$.
$\xi_A$ ($A=A^L,A^R,B$) are the gauge fixing parameters.
We take the unitary gauge, $\xi_A = \infty$, and concentrate ourselves only on the physical 
components, i.e.,  $A_\mu^L$, $A_\mu^R$ and $B_\mu$.

The boundary conditions of gauge fields at $z=R$ are given by
\begin{eqnarray}
\partial_5 A_\mu^{La}  &=& 0,
\quad a = 1,2,3,
\nonumber
\\
\quad A_\mu^{Ra} &=& 0,
\quad a = 1,2,
\nonumber
\\
\partial_5(g_5 B_\mu + \tildeg_5 A_\mu^{R3} ) &=& 0,
\quad
\tildeg_5 B_\mu - g_5 A_\mu^{R3} = 0.
\label{eq:bc-UV}
\end{eqnarray}
The boundary conditions at $z=R'$ are 
\begin{eqnarray}
\partial_z (A_\mu^{La} + A_\mu^{Ra}) &=& 0,
\quad
a = 1,2,3,
\nonumber
\\
\partial_5 B_\mu &=& 0,
\label{eq:bc-IR1}
\end{eqnarray}
and for $A_\mu^L - A_\mu^R$ we assign Robin boundary conditions
\begin{eqnarray}
(M + \partial_5) (A_\mu^{La} - A_\mu^{Ra}) &=& 0,
\quad
a=1,2,3,
\label{eq:bc-IR2}
\end{eqnarray}
where we have introduced a parameter $M$ with mass dimension one.
Boundary conditions \eqref{eq:bc-UV} \eqref{eq:bc-IR1} 
are same as ones in \cite{Cacciapaglia:2004jz}.
When $M\to\infty$, \eqref{eq:bc-IR2} becomes the Dirichlet b.c. $\partial_5(A_\mu^{La} - A_\mu^{Ra}) = 0$ as in the original model \cite{Cacciapaglia:2004jz}.

When $M=0$ the model has unbroken $SU(2)_L \times U(1)_Y$ ($Y = T^{R3} + B-L$) gauge symmetry.
Therefore $M$ can be related with a dynamics of the electroweak symmetry breaking, which lies  on the $z=R'$ brane.
Actually some boundary conditions in \eqref{eq:bc-UV}-\eqref{eq:bc-IR2} can be reproduced by introducing a mass term localized on each branes \cite{Gherghetta:2000qt,Barbieri:2003pr}.
Together with the surface terms, the boundary action is partly given by
\begin{eqnarray}
S_{\rm bdr} &\supset& \int d^4x 
\biggl\{ 
\frac{1}{2}
\left[\left(\frac{z}{R} \right)\calA^a_\mu \partial_5 \calA^{a\mu}
 + \left(\frac{z}{R}\right)^2 M_{IR} \calA^a_\mu \calA^{a\mu} \right]_{z=R'}
\nonumber\\&&
-\frac{1}{2}
\left[ \left(\frac{z}{R}\right)( A^{Ra}_\mu \partial_5 A^{Ra\mu} + B_\mu \partial_5 B^\mu)
 + \left(\frac{z}{R}\right)^2 M_{UV} u^\dag \calX_\mu \calX^\mu u \right]
\biggr\},
\nonumber
\\
\calA^a_\mu &\equiv& \frac{A_\mu^{La} - A_\mu^{Ra}}{\sqrt{2}},
\quad
\calX_\mu \equiv g_5 A_\mu^{Ra}T^{Ra} + \frac{1}{2}\tildeg_5 B_\mu,
\quad
u = \begin{pmatrix} 0 \\ 1 \end{pmatrix},
\label{eq:act-bnd}
\end{eqnarray}
where $M_{UV}$ and $M_{IR}$ are the mass parameters.
When we set
\begin{eqnarray}
M_{IR} &=& (R'/R)M,
\nonumber
\\
M_{UV} &\to& \infty,
\label{eq:bnd-mass}
\end{eqnarray} 
the boundary action reproduces boundary conditions \eqref{eq:bc-UV}-\eqref{eq:bc-IR2}.
We note that even $M_{IR}$ is as large as $1/R$, we have a small value of $M = \calO(1/R') \ll M_{IR}$ thanks to the suppression factor $R/R'$. 
In the $M \to \infty$ limit, the wave functions $A_\mu^{La} - A_\mu^{Ra}$ vanish 
and decouple completely with the source of the electroweak symmetry breaking on $z=R'$ brane, as the name ``Higgsless'' stands for.

In the low-energy effective four dimensional (4D) theory there are the photon, $Z$ and $W^{\pm}$ bosons,
and their Kaluza-Klein (KK) excitations.
The expansions to Kaluza-Klein modes are
given by
\begin{eqnarray}
A_\mu^{(L\pm)}(x,z) &=& \sum_{n=1}^\infty \psi^{(W)}_n(z) W^{\pm(n)}_\mu(x),
\nonumber
\\
A_\mu^{(R\pm)}(x,z) &=& \sum_{n=1}^\infty \psi^{(W)}_n(z) W^{\pm(n)}_\mu(x),
\nonumber
\\ 
A_\mu^{(L3)}(x,z) &=& \sum_{n=0}^\infty \psi^{(L3\gamma)}_n(z) \gamma^{(n)}_\mu(x)
+ \sum_{n=1}^\infty \psi^{(L3Z)}_n(z) Z^{(n)}_\mu(x),
\nonumber
\\
A_\mu^{(R3)}(x,z) &=& \sum_{n=0}^\infty \psi^{(R3\gamma)}_n(z) \gamma^{(n)}_\mu(x)
+ \sum_{n=1}^\infty \psi^{(R3Z)}_n(z) Z^{(n)}_\mu(x),
\nonumber
\\
B_\mu(x,z) &=& \sum_{n=0}^\infty \psi^{(B\gamma)}_n(z) \gamma^{(n)}_\mu(x)
+ \sum_{n=1}^\infty \psi^{(BZ)}_n(z) Z^{(n)}_\mu(x),
\label{eq:KK-towers}
\end{eqnarray}
where $\gamma_\mu^{(n)}(x)$, $Z_\mu^{(n)}(x)$ and $W_\mu^{\pm(n)}(x)$ 
are KK excited states with masses $m_n^{(\gamma)}$, $m_n^{(Z)}$ and $m_n^{(W)}$, respectively.
$W_\mu^{\pm} \equiv (W_\mu^1 \mp i W_\mu^2)/\sqrt{2}$ and so on.
$\gamma_\mu^{(0)}$, $Z_\mu^{(1)}$ and $W_\mu^{\pm(1)}$ correspond to the photon, $Z$-boson  and $W^{\pm}$ bosons in the SM, respectively.
Wave functions $\psi^{(A)}(z)$ satisfy bulk equations of motion (EOM)
\begin{eqnarray}
\left(
\partial_5^2 -  \frac{1}{z}\partial_5 + q^2 \right) \psi^{(A)} (z,q) &=& 0,
\label{eq:eom}
\end{eqnarray}
where we have assumed that solutions take the form of
$A_\mu^{(A)}(q)e^{-iqx} \psi_k^{(A)}(z)$.
Solutions of the EOM are written in the form of $\psi^{(A)} = z [C_1 J_1(q z) + C_2 Y_1(q z)]$, where $J_1$ and $Y_1$ are Bessel functions of the first kind and second kind, respectively.
Boundary conditions \eqref{eq:bc-UV}, \eqref{eq:bc-IR1} and \eqref{eq:bc-IR2} 
determine the KK masses and eigenfunctions of KK excitations except for their normalizations.
They are summarized in Appendix \ref{sec:wavefunc}. 

Just same as the original model \cite{Cacciapaglia:2004jz}, we assume that fermions are localized on the $z=R$ brane. The couplings of the fermions to the gauge bosons are read from the covariant derivatives
at $z=R$,
\begin{eqnarray}
\lefteqn{
\left.
\left(g A_\mu(x,z) + g'Y B_\mu(x,z)
\right) \right|_{z=R}}
\nonumber\\
&\supset&
\biggl(
g_5 \psi_1^{(L\pm)}(z) T^{L\pm} W_\mu^\pm(x)
+ g_5 T^{L3} \left[Z_\mu(x) \psi_1^{(L3Z)}(z) + \gamma_\mu(x)\psi_0^{(L3\gamma)}(z) \right]
\nonumber\\&&
+ \tildeg_5 Y \left[Z_\mu(x) \psi_1^{(BZ)}(z) + \gamma_\mu(x)\psi_0^{(B\gamma)}(z) \right]
\biggr)\biggr|_{z=R}
\nonumber\\
&=& g T^{L\pm} W_\mu^\pm(x) + g T^{L3} \left[Z_\mu(x) c_w + \gamma_\mu(x) s_w \right]
 + g' Y \left[- Z_\mu(x) s_w + \gamma_\mu(x) c_w \right],\label{eq:cov-dev}
\end{eqnarray}
where $c_w=\cos\theta_W$, $s_w = \sin\theta_W$ and $\theta_W$ is the Weinberg angle.
$g$ and $g'$ are the 4D couplings of $SU(2)_L$ and $U(1)_Y$, respectively.
In the last line of \eqref{eq:cov-dev} the couplings to the photon is given by $eQ$ where $Q = T^{L3}+Y$ is the electric charge and $e = g/\sin\theta_W$ is the electromagnetic coupling constant. Hence we obtain normalization conditions at $z=R$ as follows
\begin{eqnarray}
g_5 \psi_1^{(L\pm)}(R) &=& g,
\nonumber
\\
g_5 \psi_1^{(L3Z)}(R) &=& g\cos\theta_W,
\nonumber
\\
\tildeg_5 \psi_1^{(BZ)}(R) &=& - g'\sin\theta_W,
\label{eq:UV-bc}
\end{eqnarray}
and photon wave functions (given in \eqref{eq:photon-func}) are fixed by
\begin{eqnarray}
g_5 \psi_0^{(L3\gamma)}(z)
= 
g_5 \psi_0^{(R3\gamma)}(z)
=
\tildeg_5 \psi_0^{(B\gamma)}(z)
= e.
\label{eq:photon-bc}
\end{eqnarray}

Here let us relate 5D and 4D couplings.
From the boundary conditions \eqref{eq:UV-bc} and wave functions given in \eqref{eq:Z-func},
we obtain
\begin{eqnarray}
\frac{g'^2}{g^2} &=& -\frac{\tildeg_5}{g_5} \frac{\psi_1^{(BZ)}(R)}{\psi_1^{(L3Z)}(R)}
= \frac{\tildeg_5^2}{g_5^2 + \tildeg_5^2}.
\label{eq:rel-UV}
\end{eqnarray}
The wave-function normalization of the photon is given by
\begin{eqnarray}
Z_\gamma 
&\equiv& \int_{R}^{R'} dz \frac{R}{z} \left\{
\left(\psi_0^{(B\gamma)}(z)\right)^2 + \left(\psi_0^{(L3\gamma)}(z)\right)^2 + \left(\psi_0^{(R3\gamma)}(z) \right)^2
\right\}
\nonumber
\\
&=& R \ln(R'/R) \left(\frac{1}{\tildeg_5^2} + \frac{2}{g_5^2} \right) e^2 = 1.
\label{eq:photon-norm}
\end{eqnarray}
From \eqref{eq:rel-UV} and \eqref{eq:photon-norm}, we obtain relations between 4D and 5D gauge couplings as
\begin{align}
g^2 &= \frac{g_5^2}{R\ln(R'/R)},
& g'^2 &=\frac{g_5^2 \tildeg_5^2}{(g_5^2 + \tildeg_5^2) R \ln(R'/R)},
\end{align}
and $\sin^2\theta_W$ is given by
\begin{eqnarray}
\sin^2\theta_W &\equiv& \frac{g'^2}{g^2+g'^2}
= \frac{\tildeg_5^2}{g_5^2 + 2\tildeg_5^2}.
\label{eq:sin2W}
\end{eqnarray}
With these relations one also finds that the normalized wave functions satisfy
\begin{eqnarray}
\left. \left|\Psi^{(W)}\right|^2 \right|_{z=R}
&=&
\left. \left| \Psi^{(Z)} \right|^2 \right|_{z=R}
= \frac{1}{R\ln(R'/R)},
\end{eqnarray}
where
\begin{eqnarray}
\left| \Psi^{(Z)}\right|^2
&\equiv&
\left| \psi_1^{(L3Z)}\right|^2 + \left| \psi_1^{(R3Z)}\right|^2 + \left| \psi_1^{(BZ)}\right|^2,
\nonumber
\\
\left|\Psi^{(W)} \right|^2 
&\equiv&
\left| \psi_1^{(L\pm)}\right|^2 + \left| \psi_1^{(R\pm)}\right|^2.
\end{eqnarray}
For later use,
we define wave function renormalizations of $W$ and $Z$ bosons by
\begin{eqnarray}
Z_W &\equiv& \int_{R}^{R'} dz \frac{R}{z} \left|\Psi^{(W)}(z)\right|^2,
\nonumber
\\
Z_Z &\equiv& \int_{R}^{R'} dz \frac{R}{z} \left| \Psi^{(Z)}(z) \right|^2.
\label{eq:wave-ren}
\end{eqnarray}

Masses of $W$ and $Z$ bosons, $M_W$, $M_Z$ correspond to $m_1^{(W)}$, $m_1^{(Z)}$ and those are determined by KK mass conditions \eqref{eq:cond-W} and \eqref{eq:cond-Z}, respectively.
For $m_1^{(V)} \ll 1/R'$ ($V = W,Z$), the KK mass conditions are approximately written as
\begin{eqnarray}
\left(m_1^{(V)}\right)^2 &\simeq& \frac{x^2}{R'^2 (1 + \frac{2}{MR'})\ln(R'/R)}
\left(1 + \frac{3}{8} \frac{x^2}{(1 + \frac{2}{MR'})\ln(R'/R)}\right),
\label{eq:WZ-mass}
\end{eqnarray}
where $x^2 = 1$ [ $(g_5^2 + 2\tildeg_5^2)/(g_5^2 + \tildeg_5^2) = 1/\cos^2\theta_W$ ] for 
$V=W$ [ $Z$ ].
In the $M R'\to \infty$ limit they agree with results in \cite{Cacciapaglia:2004jz}, $m_1^{(W)} \simeq 1/(R'\sqrt{\ln(R'/R)})$ and $m_{1}^{(Z)} \simeq 1/(R'\sqrt{R'/R}\cos\theta_W)$.
When  $MR' \ll 1$, $(m_1^{(W,Z)})^2$ are suppressed by a factor $(1 + \frac{2}{MR'})^{-1} \simeq MR'/2$.

A condition $m_1^{(V)} = M_V$ essentially normalizes $R$ and $R'$, i.e., the size and shape of the extra dimension, and also determine the shapes of wave functions $\psi_n^{(A)}$.
Contrary, one can read masses of $W$ and $Z$ bosons from the bulk and boundary actions.
In the boundary action the mass terms at $z=R'$ serve masses for the $W$ and $Z$ bosons.
Such brane masses for $W$ and $Z$ bosons $m_{\text{brane}}^{(V)}$ ($V=W,Z$) can be read from the boundary interaction 
\eqref{eq:act-bnd} as
\begin{eqnarray}
\left(m^{(W)}_{\text{brane}}\right)^2 W_\mu^- W^{+\mu}(x)
&=&
\frac{R'}{R} M \left. \left( \frac{\psi_1^{(L\pm)} - \psi_1^{(R\pm)}}{ \sqrt{2} }\right)^2 \right|_{z=R'} W_\mu^- W^{+\mu} (x),
\nonumber
\\
\frac{1}{2} \left(m^{(Z)}_{\text{brane}}\right)^2 Z_\mu Z^{\mu}(x)
&=&
\frac{1}{2}\frac{R'}{R} M \left.\left( \frac{\psi_1^{(L3Z)} - \psi_1^{(R3Z)}}{ \sqrt{2} }\right)^2
\right|_{z=R'} Z_\mu Z^{\mu}(x).
\label{eq:brane-mass}
\end{eqnarray}
Using wave functions in Appendix~\ref{sec:wavefunc}, we obtain
\begin{eqnarray}
\frac{m^{(V)}_{\text{brane}}}{m_1^{(V)}} &\simeq& \sqrt{\frac{2}{MR' (1 + \frac{2}{MR'})}},
\quad V = Z,\,W,
\label{eq:brane-ratio}
\end{eqnarray}
where $C'(R,m_1^{(V)}) \simeq (m_1^{(V)})^2 R \ln(R'/R)$ has been used.
In the $MR'\to0$ limit, $m^{(V)}_{\text{brane}} = m_1^{(V)}$ ($V=Z,W$) is satisfied and hence
brane masses account for masses of the $W$ and $Z$ bosons.
When $MR' \to \infty$, on the other hand, $m_{\text{brane}}^{(V=W,Z)}$ vanish.
Alternatively, one can define
\begin{eqnarray}
P_W &\equiv& \int_{R}^{R'} dz \frac{R}{z}
\left[ 
\left(\partial_5\psi_1^{(L\pm)}\right)^2
+\left(\partial_5\psi_1^{(R\pm)}\right)^2
\right],
\nonumber
\\
P_Z
&\equiv& \int_{R}^{R'} dz \frac{R}{z}
\left[
\left(\partial_5\psi_1^{(L3Z)}\right)^2
+\left(\partial_5\psi_1^{(R3Z)}\right)^2
+\left(\partial_5\psi_1^{(BZ)}\right)^2
\right],
\label{eq:PZPW}
\end{eqnarray}
each of which measures the contribution of extra-dimensional component of the momentum, $p_5$,  to the mass-squared of the vector boson.
Contrary to the brane masses \eqref{eq:brane-mass}, in the $MR'\to \infty$ limit we obtain \cite{Cacciapaglia:2004jz}
\begin{eqnarray}
(P_W,\,P_Z) &\stackrel{MR'\to\infty}{=}& (M_W,M_Z).
\label{eq:PWZ-limit}
\end{eqnarray}
Hence in this limit $P_{W}$ and $P_{Z}$ account for the $W$ and $Z$ boson masses $M_W$ and $M_Z$, respectively.

Now we consider the precision observables.
The $S$, $T$ and $U$ parameters are defined in \cite{Peskin:1990zt,Peskin:1991sw} by
\begin{eqnarray}
S &\equiv& 16\pi [\Pi'_{33}(0)- \Pi'_{3Q}(0)] ,
\nonumber
\\
T &\equiv& \frac{4\pi}{c_w^2 s_w^2 M_Z^2} \left[ \Pi_{11}(0) - \Pi_{33}(0) \right],
\nonumber
\\
U &\equiv& 16\pi [\Pi'_{11}(0) - \Pi'_{33}(0)].
\end{eqnarray}

Since $S$ and $U$ parameters are related with wave-function renormalizations \cite{Kennedy:1988sn,Morii:2004tp},
just following \cite{Cacciapaglia:2004jz}, we write
\begin{eqnarray}
S &=& 16\pi \Pi'_{33}(0) = 16 \pi \frac{1-Z_Z}{g^2 + g'^2},
\nonumber
\\
U &=& 16\pi \left[\frac{1-Z_W}{g^2} - \frac{1-Z_Z}{g^2+g'^2}\right],
\end{eqnarray}
where $Z_W$ and $Z_Z$ are defined in \eqref{eq:wave-ren}, and 
we have used $\Pi_{3Q}^{(\prime)}=0$ at tree level.

There are a few possible expressions of the $T$ parameter.
At first, following \cite{Cacciapaglia:2004jz} one can identify $P_W$ and $P_Z$ with
vacuum polarizations at zero momentum
\begin{eqnarray}
P_W &\Leftrightarrow& g^2 \Pi_{11}(0),
\nonumber
\\
P_Z &\Leftrightarrow& (g^2 + g'^2) \Pi_{33}(0),
\label{eq:identify}
\end{eqnarray}
and define
\begin{eqnarray}
T = T_{(\rmA)} 
&\equiv& 
\frac{1}{\alpha_{EM} M_Z^2} \left[ \frac{P_W}{\cos^2\theta_W} - P_Z \right].
\end{eqnarray}
We note that in the $MR' \to \infty$ limit we have
\eqref{eq:PWZ-limit}
and hence identifications 
\eqref{eq:identify} are naturally allowed.
For $MR' \ll 1$, however, both $P_W$ and $P_Z$ can be much smaller than $M_W^2$ and $M_Z^2$ and the above identifications cannot be justified.
As one of alternatives to $T_{(\rmA)}$,
we express $T$ parameter by a deviation of a tree-level $\rho$ parameter from the unity.
To make a contrast with $T_{(\rmA)}$,
here we write the $\rho$ parameter in terms of $m_1^{(W)}$ and $m_1^{(Z)}$ as
\begin{eqnarray}
\rho &\equiv& \frac{1}{\cos^2\theta_W}
\left( \frac{m_1^{(W)}}{m_1^{(Z)}} \right)^2,
\end{eqnarray}
where $m_1^{(W)}$ and $m_1^{(Z)}$ are determined by the KK conditions \eqref{eq:cond-W} and \eqref{eq:cond-Z} with couplings satisfying \eqref{eq:sin2W},
respectively. Then we define
\begin{eqnarray}
T = T_{(\rmB)} &\equiv& \frac{\sin^2 \theta_W - \sin^2\theta'_W}{\alpha_{EM} \cos^2\theta_W},
\nonumber
\\
\sin^2\theta'_W &\equiv& 1 - \left(\frac{{m_1^{(W)}}}{ {m_1^{(Z)}}}\right)^2.
\end{eqnarray}
Using \eqref{eq:WZ-mass}, we estimate
\begin{eqnarray}
T_{(\rmB)} &\simeq& \frac{\sin^2\theta_W}{\alpha_{EM}\cos^2\theta_W}\cdot
\frac{3}{8(1 + \frac{2}{MR'})\ln(R'/R)}
\nonumber
\\
&\sim& 0.5 \cdot \frac{30}{(1 + \frac{2}{MR'})\ln(R'/R)}.
\end{eqnarray}
In the $MR'\to\infty$ limit we obtain $T_{(\rmB)} \sim 0.5\cdot 30/\ln(R'/R)$.
This is a considerably large value even for $\ln(R'/R) = \calO(30)$, although $m_{1}^{(W,Z)}$ are not directly related to masses of $W$ and $Z$ bosons.
For $MR' \ll 1$, on the other hand, $T_{(\rmB)}$ is suppressed by the factor $MR'$.
We also note that in the $MR'\to\infty$ limit we have $S = 6\pi/(g^2\ln(R'/R))$ \cite{Cacciapaglia:2004jz} hence
\begin{eqnarray}
S &=& 4\cos^2\theta_W \cdot T_{(\rmB)}
\label{eq:STB}
\end{eqnarray}
is satisfied.
In the followings, we use both $T_{(\rmA)}$ and $T_{(\rmB)}$ as reference values of the $T$ parameter in this model.

\section{Numerical study}\label{sec:numeric}
In the numerical study, to see the tree level effects we use
$\alpha_{EM} = e^2/4\pi = 1/128$,
$\cos\theta_W = M_W/M_Z$,
$M_W = 80.4\GeV$ and 
$M_Z = 91.2\GeV$.
We choose $R$ and $MR'$ as input parameters. $R'$ is normalized so that $m_1^{(Z)} = M_Z$ is satisfied.

\begin{figure}[tbp]
\centerline{\includegraphics[width=12cm]{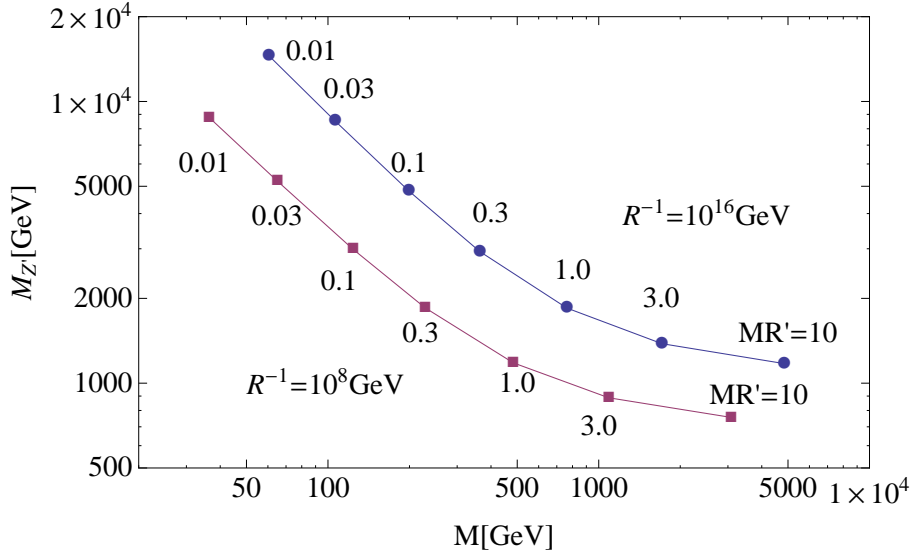}}
\caption{$(M,M_{Z'})$ as functions of $MR'$. Blue circles and red squares
correspond to $R^{-1} = 10^{16}\GeV$ and $10^8\GeV$, respectively.}\label{fig:MMZ}
\end{figure}

\begin{figure}[tbp]
\centerline{\includegraphics[width=12cm]{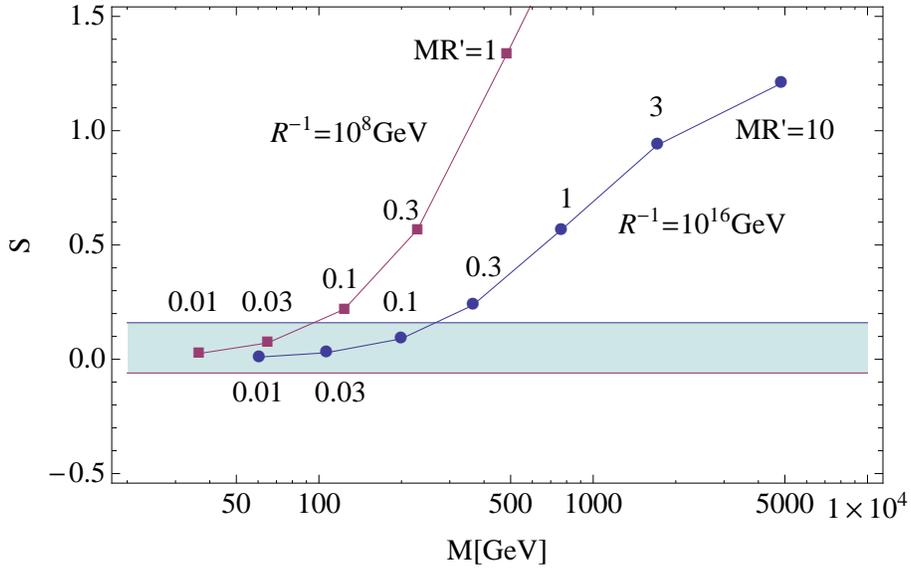}}
\caption{$(M,S)$ as functions of $MR'$. Blue circles and red squares
correspond to $R^{-1} = 10^{16}\GeV$ and $10^8\GeV$, respectively.
The light-blue horizontal band shows allowed range of the $S$ parameter.}\label{fig:MS}
\end{figure}

In Table~\ref{tbl:results}, we have tabulated  $M$, and
masses of the first KK $Z'$, $W'$, $\gamma'$.
Here $Z',\,W'$ and $\gamma'$ are first KK $Z$, $W$, and photon 
and correspond to
$Z_\mu^{(2)},\,W_\mu^{(2)}$ and $\gamma_\mu^{(1)}$ in \eqref{eq:KK-towers}, respectively. 
We also note that masses of $W'$ and $Z'$ are almost degenerate.
We have also shown $P_Z$ and $P_W$, which are defined in \eqref{eq:PZPW}.
Finally, in Table~\ref{tbl:results}, we have tabulated $S$, $T=T_{(\rmA)},\,T_{(\rmB)}$ and $U$ parameters.
\begin{table}[tbp]
\caption{%
Boundary condition parameter $M$,
masses of the first KK states $M_{Z',W',\gamma'}$,
KK momentum mass-squared $P_{Z,W}$,
and oblique parameters $S$, $T$ and $U$.
$M_{Z'}$, $M_{W'}$ and $M_{\gamma'}$ are masses of 1st KK $Z$, $W$ and photon, respectively.
$P_{Z,W}$ are KK momentum mass-squared of $Z$ and $W$ bosons (see text).
For the $T$ parameter, two different values $T=T_{(\rmA)}$ and $T_{(\rmB)}$ are shown (see text).
As input parameters, $R^{-1} = 10^{16}\GeV$, $10^8\GeV$ and $MR'\ge 0.01$ are chosen.
}\label{tbl:results}
\begin{center}
\begin{tabular}{l|ccccc|ccccc}
\hline
 & \multicolumn{5}{c|}{$R^{-1} = 10^{16}\GeV$} & \multicolumn{5}{c}{$R^{-1} = 10^{8}\GeV$} 
\\
$MR'$ & $\infty$ &$10$ & $1$ & $0.1$ &$0.01$ & $\infty$ & $10$ & $1$ & $0.1$ & $0.01$  
\\
\hline
$M$ [GeV] 
& $\infty$ & $4843$ & $763$ & $199$ & $60.5$ 
& $\infty$ & $3085$ & $483$ & $124$ & $36.4$ 
\\
$M_{Z',W'}$ [TeV]
& $1.07$ & $1.17$ & $1.85$ & $4.81$ & $14.5$
& $0.69$ & $0.76$ & $1.18$ & $3.00$ & $8.77$  
\\
$M_{\gamma'}$ [TeV]
& $1.09$ & $1.19$ & $1.87$ & $4.90$ & $14.9$
& $0.71$ & $0.78$ & $1.23$ & $3.16$ & $9.36$
\\
$P_Z$ [$\GeV^2$]
& $8190$ & $6824$ & $2723$ & $387.7$ & $40.5$
& $8022$ & $6680$ & $2625$ & $375.0$ & $38.9$ 
\\
$P_W$ [$\GeV^2$]
& $6365$ & $5306$ & $2122$ & $302.7$ & $31.6$
& $6234$ & $5197$ & $2075$ & $294.9$ & $30.7$
\\
$S$ 
& $1.36$ & $1.21$ & $0.57$ & $0.09$ & $0.01$
& $3.15$ & $2.80$ & $1.33$ & $0.22$ & $0.02$
\\
$T_{(\rmA)}$ 
& $-0.002$ &$0.05$ & $0.11$ & $0.03$ & $0.003$
& $-0.014$ &$0.11$ & $0.27$ & $0.07$ & $0.01$
\\
$T_{(\rmB)}$ 
& $0.45$ & $0.40$ & $0.19$ & $0.03$ & $0.003$ 
& $1.10$ & $0.97$ & $0.45$ & $0.07$ & $0.01$
\\
$U\times10^4$ 
& $-61$ & $-47$ & $-10$ & $-0.26$ & $-0.003$
& $-347$ & $-270$ & $-60$ & $-1.6$ & $-0.02$
\\
\hline
\end{tabular}
\end{center}
\end{table}
We also plotted the $(M,M_{Z'})$ and $(M,S)$ with respect to $MR'$ in
Figures~\ref{fig:MMZ} and \ref{fig:MS}, respectively.

From Table~\ref{tbl:results} one finds that
\begin{eqnarray}
M_{Z'},\,M_{W'} \simeq 2.4/R',
\label{eq:MZprime}
\end{eqnarray}
and that $\gamma'$ is slightly heavier than $Z'$ and $W'$. 
From plots in Figure~\ref{fig:MMZ}, we see that $M_{Z'}$ (or $R'$) is in inverse proportion to $M$ when $M \lesssim 300\GeV$.
For $MR' \ll 1$ one finds an approximation
\begin{eqnarray}
R'^{-1} &\simeq& \frac{2M_W^2}{M} \ln[R^{-1}/\mu']
\nonumber
\\
&=& 3.9\TeV \cdot \left( \frac{100\GeV}{M}\right) 
\left[
1 + \frac{1}{30} \left\{
\ln\left(\frac{R^{-1}}{10^{16}\GeV}\right) - \ln\left(\frac{\mu'}{10^3\GeV}\right)
\right\}
\right],
\label{eq:MZM}
\end{eqnarray}
where $\mu' = \calO(1\TeV)$, or one can solve $\mu' = R'^{-1}$ by an iteration.

Experimental lower bound for masses of heavy charged vector bosons at LHC Run-1 are,
$M_{W'} \ge 3.24\TeV$ at 95\% CL \cite{ATLAS:2014wra} for the sequencial standard model (SSM), and $M_{W'} \ge 2.7\TeV$ at 95\% CL \cite{Khachatryan:2015pua} for universal fermion couplings. For neutral vector bosons, we have
$M_{Z'} \ge 2.79\TeV$ at 95\% CL \cite{Aad:2014cka} for $Z'$ with SM-like coupling to fermions, and $M_{Z'} \ge 2.90$ at 95\% CL \cite{Khachatryan:2014fba} for SSM.
From the experiments at LHC Run 2 ($\sqrt{s} = 13\TeV$), 
similar or slightly stringent bounds are obtained \cite{1512.01224,CMS15W,CMS15Wtb,CMS15Z,ATLAS15W,ATLAS15Z}.
Hence we safely put the experimental bounds as $M_{Z'},\,M_{W'} \gtrsim 3\TeV$, 
and obtain bounds $MR' \lesssim 0.3$ [$0.1$] for $R^{-1} = 10^{16}\GeV$ [$10^8\GeV$].

For $P_Z$ and $P_W$, one finds numerically that 
\begin{eqnarray}
P_V \simeq M_V^2 \left(1 + \frac{2}{MR'}\right)^{-1},
\quad
V = W,\,Z,
\label{eq:PA}
\end{eqnarray}
and the correspondence \eqref{eq:identify} holds only when $MR' \gg 1$.

For the $S$ parameter, as pointed out in \cite{Cacciapaglia:2004jz} in the $M\to\infty$ limit large value of $S = \calO(1)$ is obtained.  
We also see that $S$ shrinks as $MR'$ decreases.
In Figure~\ref{fig:MS}, an allowed region of $S$ parameter is also shown.
Here current experimental bounds for $S,T,U$ are given in \cite{Baak:2014ora} as
\begin{eqnarray}
S = 0.05 \pm 0.11,
\quad
T = 0.09\pm 0.13,
\quad
U = 0.01\pm 0.11,
\end{eqnarray}
and $S-T$, $S-U$ and $T-U$ correlations are $0.90$, $-0.59$ and $0.83$, respectively. 
From the allowed range in Figure.~\ref{fig:MS} we obtain the bound $MR' \lesssim 0.2$ [$0.05$] for $R^{=1} = 10^{16}\GeV$ [$10^8\GeV$].

For the $T$ parameter,
$T=T_{(\rmA)}$ is tiny for $MR'=\infty$ and this is consistent with the results in 
\cite{Cacciapaglia:2004jz}.
When $MR'$ decreases, $MR'$ monotonically increases [decreases] for $\infty > MR' \gtrsim 1$ [$1 \gtrsim MR > 0$].

$T_{(\rmB)}$ is monotonically decreasing for decreasing $MR'$, and 
one finds that $T_{(\rmB)}$ is almost proportional to $S$,
\begin{eqnarray}
3 T_{(\rmB)} &\sim& S,
\label{eq:TB-fit}
\end{eqnarray}
from which we find that \eqref{eq:STB} is well satisfied for finite $MR'$ since $4\cos^2\theta_W = 3.1$.
Numerically we also find that
\begin{eqnarray}
T_{(\rmA)} &\simeq& T_{(\rmB)},
\label{eq:TA-TB}
\end{eqnarray}
for $M_{Z'} \gtrsim 3\TeV$.
$U$ parameter is very small and this also agrees with results in the original model \cite{Cacciapaglia:2004jz}.

As we have seen from  Table~\ref{tbl:results}, Figures~\ref{fig:MMZ} and \ref{fig:MS},
both $M_{Z'}$ and oblique parameters depend largely on both $R$ and $MR'$. 
However, once we choose the free parameters as $M_{Z'}$ and $R$, 
we find that the oblique parameters mainly depend on $M_{Z'}$ but weakly on $R$. 
We numerically find that $S$ and $M_{Z'}$ are related by
\begin{eqnarray}
S &\simeq& \left(\frac{1350\GeV}{ M_{Z'}}\right)^{1.92},
\label{eq:S-fit}
\end{eqnarray}
irrespective to the magnitude of $R$. 
This behavior is reasonably reflects the fact that 
$S$ is a dimension-six operator and should be inversely proportional to the square of a new physics scale.

\begin{figure}[tbp]
\centerline{\includegraphics[width=12cm]{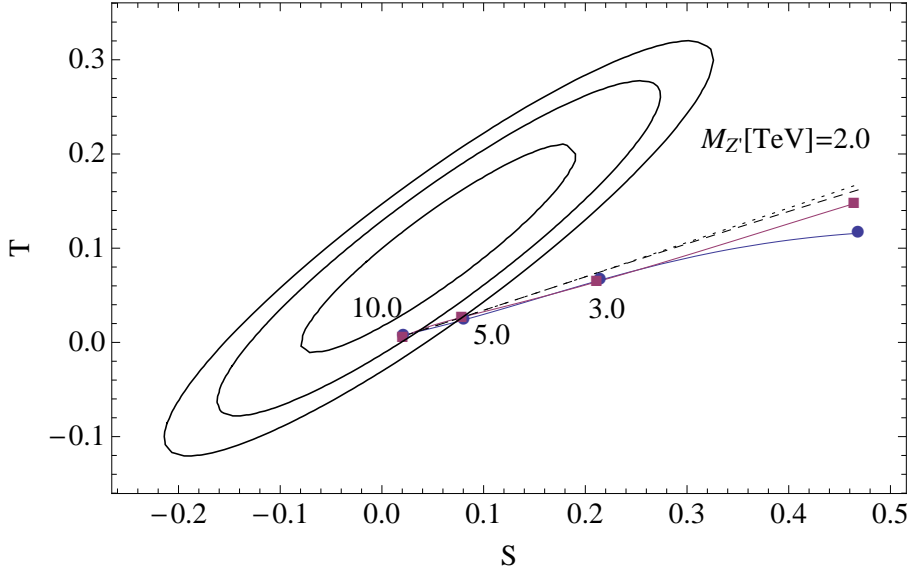}}
\caption{%
$(S,T)$ as functions of the first KK $Z$ boson mass $M_{Z'}$, $M_{Z'} = [2\TeV, 10\TeV]$.
The origin $(S,\,T)=(0,\,0)$ corresponds to the SM value.
Blue circles and red squares correspond to $(S,\,T=T_{(\rmA)})$ for $R^{-1}=10^{16}\GeV$ and $10^8\GeV$, respectively.
Black dashed and dotted lines indicate $(S,\, T=T_{(\rmB)})$ for $R^{-1} = 10^{16}\GeV$ and $10^{8}\GeV$, respectively.
Contours are $68\%$, $95\%$ and $99\%$ CL in $(S,\,T)$ plane with $U=0$.
All points and curves meet at $(S,T)=(0,0)$ for $M_{Z'}=\infty$. Plots for $M_{Z'}> 10\TeV$ are busy and omitted.}
\label{fig:STMZ}
\end{figure}

From \eqref{eq:TB-fit}, \eqref{eq:TA-TB} and \eqref{eq:S-fit},
for $M_{Z'} \gtrsim 3\TeV$ one can write both $S$ and $T$ as functions of $M_{Z'}$ irrespective to the magnitude of $R$.
In Fig.~\ref{fig:STMZ}, we plot $(S,\,T)$ with respect to $M_{Z'}$.
From the constraints for $(S,T)$ with $U\simeq0$ which is shown in Figure~\ref{fig:STMZ}, we obtain 
\begin{eqnarray}
M_{Z'} \ge 5\TeV \, \text{[$8\TeV$] at 99\% [95\%] CL.}
\label{eq:MZbnd}
\end{eqnarray}
This result is almost irrespective to $R$.
Hereafter we refer \eqref{eq:MZbnd} as $95\%$ and $99\%$ CL bounds of this model.

From the Fig.~\ref{fig:MMZ} (or \eqref{eq:MZM}) and the bounds \eqref{eq:MZbnd},
we obtain 
\begin{eqnarray}
M &\lesssim& 
\begin{cases}
200\GeV \, [120\GeV] &\text{for $R^{-1} = 10^{16}\GeV$}
\\
70\GeV \, [40\GeV]  &\text{for $R^{-1} = 10^8\GeV$}
\end{cases}
\end{eqnarray}
and
\begin{eqnarray}
MR' &\lesssim&
\begin{cases}
0.1 \, [0.03] &\text{for $R^{-1} = 10^{16}\GeV$}
\\
0.03 \, [0.01] &\text{for $R^{-1} = 10^8\GeV$}
\end{cases}
\end{eqnarray}
for 99\% [95\%] CL bounds.

If we assume that the boundary condition \eqref{eq:bc-IR2} comes from the boundary mass terms given in \eqref{eq:act-bnd} with \eqref{eq:bnd-mass}, then from Eq.~\eqref{eq:brane-ratio}, we obtain
\begin{eqnarray}
\frac{m^{(V)}_{\text{brane}}}{M_V}
\simeq
\frac{m^{(V)}_{\text{brane}}}{m_1^{(V)}}
&\gtrsim&
\begin{cases}
 0.976 \, [0.993] & \text{for $R^{-1} = 10^{16}\GeV$}
\\
 0.993 \, [0.998] & \text{for $R^{-1} = 10^{8}\GeV$}
\end{cases}
\nonumber
\\
&& (V = W,Z),
\end{eqnarray}
for $99\%$ [95\%] CL bounds. Since $R^{-1}$ cannot exceed the reduced Planck mass $\overline{M}_{\rm Pl}$, by extrapolating the above results up to $R^{-1} = \overline{M}_{\rm Pl} = 2.4\times10^{18}\GeV$ we find that the brane mass terms account for more than $97\%$ [$99\%$] of $W$ and $Z$ boson masses for $99\%$ [$95\%$] CL bounds with $R^{-1} \le \overline{M}_{\rm Pl}$.

The boundary mass term at $z=R'$ may be interpreted as a VEV of a scalar field $\Phi$, which is a scalar transforming as $(\bm{2},\bar{\bm{2}})$ of $SU(2)_L \otimes SU(2)_R$ and is localized on the $z=R'$ brane.
It is natural to identify this scalar with the SM-like Higgs field with $125\GeV$ mass.
If so, the ratios \eqref{eq:brane-ratio} are viewed as a ratios of the
$HWW$ and $ZWW$ couplings to their SM values, i.e., $\kappa_V \equiv g_{HVV}/g_{HVV}^{SM}
\simeq (m_{\text{brane}}^{(V)}/m_1^{(V)})^2$ ($V=W,Z$)
where $g_{HVV}^{SM} = g M_V^2$ are $HVV$ couplings in the SM.
Then one obtain $\kappa_W = \kappa_Z$ and
\begin{eqnarray}
1-\kappa_{W,Z}  &\simeq& \frac{MR'}{2 + MR'}
\simeq M_W^2 R'^2 \ln(R'/R)
\nonumber\\
&=& \left(\frac{2.4M_W}{M_{Z'}}\right)^2 \ln\left(\frac{2.4}{M_{Z'}R}\right),
\end{eqnarray} 
where \eqref{eq:WZ-mass} with $M_W \simeq m_1^{(W)}$ and \eqref{eq:MZprime} are used.
For bounds $1-\kappa_V \le 10\%$, $5\%$, $1\%$ and $0.5\%$,
we have constraints $MR' \le 0.22$, $0.105$, $0.020$ and $0.010$, respectively.
$M_{Z'}\ge 3\TeV$, $5\TeV$, $10\TeV$ and $15\TeV$ with $R = 10^{16}\GeV$ [$10^8\GeV$] correspond to
$1-\kappa_{W,Z} \le 12\%$, $4\%$, $1\%$ and $0.5\%$ [$5\%$, $1.6\%$, $0.4\%$ and $0.2\%$], respectively.
$\kappa_{V=W,Z}$ will be precisely measured at current and future collider experiments.
Hence both the mass of first KK bosons and the couplings between the Higgs and weak bosons will constrain the parameters of the model.

\section{Summary and Discussion}\label{sec:summary}
In this paper we reconsidered the Higgsless model in the warped extra dimension.
Some Dirichlet boundary conditions on the TeV brane are replaced with 
Robin boundary conditions which are parameterized by a mass parameter $M$.
The Peskin-Takeuchi oblique parameters in this model at tree level are evaluated.
From the experimental bounds of oblique parameters the lower bounds of the mass of the first Kaluza-Klein excited $Z$ and $W$ bosons $M_{Z',W'}$ are obtained. At 95 \% [99 \%] confidence level (CL), $M_{Z'}$, $M_{W'}$ are greater than $8 \TeV$ [$5 \TeV$]. 
The magnitude of $M$, which is infinity in the original model, is smaller than 120 [40] GeV for the curvature of the warped space $R^{-1}=10^{16} \GeV$ [$10^{8} \GeV$]
at 95\% CL. If we assume that the Robin boundary conditions come from the brane mass terms,
it turns out that the brane mass accounts for more than 97\% of the W and Z boson masses for $99\%$ CL bounds.
If the brane mass is induced by the vacuum expectations value of the Higgs field $\Phi$ localized on the TeV brane, the model will also be tested by the precision measurement of the Higgs-weak boson couplings.

In this model fermions corresponding to the SM right-handed fermions have not been introduced.
To obtain the Yukawa coupling, at least either left-handed fermions or Higgs field $\Phi$, or both must propagate in the bulk.
When $\Phi$ propagates in the $AdS_5$ bulk and its kinetic term is given by $\int d^4x \int dz (R/z)^3 \eta^{MN}\tr (D_M \Phi)^\dag(D_N \Phi)$,
then a steeply growing VEV, $\langle \Phi(z) \rangle\propto z^\alpha$, $\alpha > 1$,
eaasily mimics the boundary mass term at $z=R'$ in \eqref{eq:act-bnd}.
We also note that the hierarchy between $M$ and $R'^{-1}$ which is expressed as $MR'$ can be ameliorated to $\calO( (MR')^{1/\alpha})$.
In the $\alpha=3$ case, $\langle \Phi(z)\rangle$ can be viewed as a condensation which breaks $SU(2)_R\times SU(2)_R$ ``chiral'' symmetry in AdS/QCD \cite{ArkaniHamed:2000ds,Erlich:2005qh,DaRold:2005vr} in the context of $AdS$/CFT correspondence \cite{Maldacena:1997re,Witten:1998qj}.
In the $\alpha=2$ case, $\langle \Phi(z)\rangle$ may be interpreted as a VEV of 5th component of the $SO(5)/SO(4)$ gauge fields in the context of the gauge-Higgs unification (GHU) in warped space \cite{Hosotani:2008tx,Hosotani:2010hx,Funatsu:2013ni,GHU-Funatsu,GHGU},
or as a pseudo Nambu-Goldstone boson of $SO(5) \to SO(4)$ symmetry breaking \cite{Agashe:2004rs,Contino:2010rs,Gherghetta:2010cj,Csaki2015}.
In the GHU case, the electroweak symmetry will be broken by the Hosotani mechanism \cite{Hosotani:1983xw}, and the mass of the Higgs is stabilized by the higher-dimensional gauge symmetry \cite{Hatanaka:1998yp}.

In this paper contributions to oblique parameters at loop levels are not evaluated.
In this model the mass and the mechanism to develop a VEV of the ``Higgs'' are also unexplained. These issues are model-dependent and will be discussed separately.

\subsection*{Acknowledgements}
This work was supported in part by National Research Fund of Korea\\
(Grant No. 2012R1A2A1A01006053).
\appendix
\section{Wave functions}\label{sec:wavefunc}
\subsection{Bulk functions}
It would be useful to introduce bulk functions 
$C(z,q),S(z,q)$ which satisfy the equation of motion \eqref{eq:eom}
and satisfy
\begin{align}
C(R',q) &= 1,
& S(R',q) &= 0,
\nonumber
\\
C'(R',q) &= 0,
&
S'(R',q) &= q, 
\end{align}
where $C'(z,q) \equiv \partial_5 C(z,q)$ and $S'(z,q) \equiv \partial_5 S(z,q)$.
They can be written by
\begin{eqnarray}
C(z,q) &=& \frac{\pi}{2}qz [Y_0(qR')J_1(qz) - J_0(qR')Y_1(qz)],
\nonumber
\\
S(z,q) &=& \frac{\pi}{2}qz [-Y_1(qR')J_1(qz) + J_1(qR')Y_1(qz)],
\nonumber
\\
C'(z,q) &=& \frac{\pi}{2}q^2z [Y_0(qR')J_0(qz) - J_0(qR')Y_0(qz)],
\nonumber
\\
S'(z,q) &=& \frac{\pi}{2}q^2z [-Y_1(qR')J_0(qz)+J_1(qR')Y_0(qz)],
\end{eqnarray}
where $J_\nu$ and $Y_\nu$ are Bessel functions of 1st and 2nd kind, respectively.
$C,S,C'$ and $S'$ satifty
\begin{eqnarray}
C(z,q)S'(z,q) - C'(z,q) S(z,q) &=& \frac{qR}{R'}.\label{eq:bulk-rel}
\end{eqnarray}
From the boundary conditions at $z=R'$, Eqs.~\eqref{eq:bc-IR1}\eqref{eq:bc-IR2}, one can write the wave functions in \eqref{eq:KK-towers} as
\begin{eqnarray}
\psi^{(LaU)} + \psi^{(RaU)} &=& a_V^{(U)} C(z,q),
\nonumber
\\
\psi^{(LaU)} - \psi^{(RaU)} &=& a_{A}^{(U)} [S(z,q) - (q/M) C(z,q)],
\quad a  = 1,2,3,
\nonumber
\\
\psi^{(BU)} &=& a_{B}^{(U)} C(z,q),
\end{eqnarray}
for $U=Z,\,\gamma,\,W$. Subscripts for the KK number are omitted.
Boundary conditions \eqref{eq:bc-UV}
determine $q = m_n^{(U)}$ and $a_{V,A,B}^{(U)}$ except for overall normalizations.

\subsection{Charged bosons}
\paragraph{$W$-boson tower}
Wave functions for the $W^{\pm}$ bosons and their KK excitation modes are
\begin{eqnarray}
\psi_n^{(R\pm)} &=& N_{W_n} \left[S'(R) C(z) + C'(R) S(z) - 2\frac{m_n^{(W)}}{M} C'(R) C(z) \right],
\nonumber
\\
\psi_n^{(L\pm)} &=& N_{W_n} \left[S'(R) C(z) - C'(R) S(z) \right],
\end{eqnarray}
where $C(z) = C(z,m_n^{(W)})$ and so on.
$N_{W_n}$ is a normalization factor. The KK mass $m_n^{(W)}$ is determined by
\begin{eqnarray}
- 2 \frac{m_n^{(W)}}{M} C C' + C S' + S C' &=& 0,
\label{eq:W-cond}
\end{eqnarray}
where $C^{(\prime)} = C^{(\prime)}(R,m_n^{(W)})$ and so on.
Using \eqref{eq:bulk-rel} we rewrite \eqref{eq:W-cond} as
\begin{eqnarray}
- 2 \frac{m_n^{(W)}}{M} C C' + 2S C' + \frac{m_n^{(W)} R}{R'} &=& 0.
\label{eq:cond-W}
\end{eqnarray}

For the $W$ bosons ($n=1$), with the normalization condition \eqref{eq:UV-bc}, the normalizeion factor $N_{W_1}$
is determined to be
\begin{eqnarray}
N_{W_1} &=& \frac{R'}{m_1^{(W)} R \sqrt{R \ln(R'/R)}}.
\label{eq:W-norm}
\end{eqnarray}

\subsection{Neutral bosons}

\paragraph{Photon tower}
For $n\ge1$ we obtain
\begin{eqnarray}
\psi_n^{(L3\gamma)}(z) = \psi_n^{(R3\gamma)}(z)  
&=& N_{\gamma_n} \tildeg_5 C(z,m_n^{(\gamma)}) ,
\nonumber
\\
\psi_n^{(B\gamma)}(z) &=& N_{\gamma_n} g_5 C(z,m_n^{(\gamma)}),
\end{eqnarray}
where $N_{\gamma_n}$ is a normalization factor. The KK mass $m_n^{(\gamma)}$ is determined by
\begin{eqnarray}
C'(R, m_n^{(\gamma)}) &=& 0.
\end{eqnarray}
The photon correspond to the $n=0$ mode and its wave functions are given by
\begin{eqnarray}
\left(\psi_0^{(R3\gamma)}(z),\, \psi_0^{(L3\gamma)}(z),\, \psi_0^{(B\gamma)}(z) \right)
&=& N_\gamma \left(\frac{1}{g_5},\, \frac{1}{g_5},\, \frac{1}{\tildeg_5} \right),
\label{eq:photon-func}
\end{eqnarray}
where $N_\gamma=e$ is fixed by \eqref{eq:photon-bc}.

\paragraph{$Z$-boson tower}
wave functions of the $Z$ boson and its KK excitations are given by
\begin{eqnarray}
\psi_n^{(BZ)}(z) &=&
 - \frac{2\tildeg_5}{g_5} N_{Z_n} \left[S'(R) - \frac{m_n^{(Z)}}{M}C'(R) \right] C(z),
\nonumber
\\
\psi_n^{(L3Z)}(z) &=& N_{Z_n} \left[S'(R) C(z) - C'(R) S(z)\right],
\nonumber
\\
\psi_n^{(R3Z)}(z)  &=& N_{Z_n} \left[
 S'(R) C(z) + C'(R) S(z) - 2\frac{m_n^{(Z)}}{M} C'(R) C(z) \right], 
\label{eq:Z-func}
\end{eqnarray}
where $C(z) = C(z,m_n^{(Z)})$, and so on, and $N_{Z_n}$ is a normalization factor.
The KK mass $m_n^{(Z)}$ is determined by
\begin{eqnarray}
- 2\frac{m_n^{(Z)}}{M} (g_5^2 + \tildeg_5^2) C C'
 + (g_5^2 + 2\tildeg_5^2) C S'  + g_5^2 S C' &=& 0, 
\label{eq:Z-cond}
\end{eqnarray}
where $C^{(\prime)} \equiv C^{(\prime)}(R,m_n^{(Z)})$ and so on.
Using \eqref{eq:bulk-rel} we rewrite \eqref{eq:Z-cond} as
\begin{eqnarray}
- 2\frac{m_n^{(Z)}}{M} C C' + 2S C' + \frac{g_5^2 + 2\tildeg_5^2}{g_5^2 + \tildeg_5^2} \frac{m_n^{(Z)} R}{R'} &=& 0.
\label{eq:cond-Z}
\end{eqnarray}

For the $Z$ boson ($n=1$), the normalization factor is determined by \eqref{eq:UV-bc}
to be
\begin{eqnarray}
N_{Z_1} &=& \frac{R'\cos\theta_W}{m_1^{(Z)}R \sqrt{R\ln(R'/R)}}.
\label{eq:Z-norm}
\end{eqnarray}

\end{document}